\documentclass[twocolumn, prl, showpacs]{revtex4}

\usepackage{amssymb}
\usepackage{graphicx}
\usepackage{dcolumn}
\usepackage{bm}
\usepackage{amsmath}

\def\OMIT#1 {{}}
\def\MEMO#1 {{}}

\begin{document}
\title{Interaction-induced anomalous  transport behavior in one dimensional optical lattice}

\author{Zi Cai$^1$}
\author{Lei Wang$^1$}
\email{wanglei@aphy.iphy.ac.cn}
\author{ X. C. Xie$^{2,1}$}
\author{Yupeng Wang$^1$}

\affiliation{$^{1}$ Beijing National Laboratory for Condensed Matter
Physics, Institute of Physics, Chinese Academy of Sciences, Beijing
100080, P. R. China}

\affiliation{$^{2}$Department of Physics, Oklahoma State University,
Stillwater, Oklahoma 74078, USA}

\date{\today }

\begin{abstract}
The non-equilibrium dynamics of spin impurity atoms in a strongly
interacting one-dimensional (1D) Bose gas under the gravity field is
studied. We show that due to the non-equilibrium preparation of the
initial state as well as the interaction between the impurity atoms
and other bosons, a counterintuitive phenomenon may emerge: the
impurity atoms could propagate upwards automatically in the gravity
field . The effects of the strength of interaction, the gradient of
the gravity field, as well as the different configurations of the
initial state are investigated by studying the time-dependent
evolution of the 1D strongly interacting bosonic system using
time-evolving block decimation (TEBD) method. A profound connection
between this counterintuitive phenomenon and the repulsive bound
pair is also revealed.
\end{abstract}


\maketitle

Recently, ultracold atoms in optical lattice have provided a perfect
platform for simulating quantum many-body models in condensed matter
physics\cite{Greiner,Jordens}.  What's more, the uniqueness of cold
atomic system, such as the low dissipation rate as well as long
coherence times,  has opened exciting possibilities for studying
non-equilibrium quantum dynamics of many-body systems. It is known
that, due to the energy dissipation between the system and the
environment, the properties of the many-body system in solid state
physics is mainly determined by its ground state as well as its
low-energy excited state. In the ultracold atomic systems, however,
the extreme low dissipation rates there guarantees the conservation
of the system energy and total particle number in a relatively long
time. Therefore, not only the ground state but also the high energy
excited state may contribute to the non-equilibrium dynamics of the
many-body systems, which may exhibit novel phenomena
\cite{Kinoshita, Rigol,Manmana,Altman,Kollath, Bruun}.

Repulsively bound atom pairs is one of the most interesting and
novel phenomena emerging from the non-equilibrium many-body physics
in optical lattice\cite{Winkler,Daley}. It is shown that though
repulsive force separates particles in free space, under a periodic
potential and in the absence of dissipation, a bound atom pairs
could  be stabilized by the strongly repulsive interaction due to
the conservation of energy. This unconventional phenomenon provides
a typical example of non-equilibrium physics determined by the
high-energy excited state of the Hamiltonian rather than the ground
state, and has no analogue in traditional condensed matter systems
due to the rapid dissipation. In this Letter, we provide another
example of these exotic non-equilibrium phenomena by studying the
propagation of  spin impurity atoms through a strongly interacting
one-dimensional (1D) Bose gas under an external potential which
decreases linearly along a particular direction. The ground state
and the quench dynamics of the Bose-Hubbard model under linear
external potential have been discussed
previously\cite{Sachdev,Sengupta}. We show that if the initial state
is prepared far away from the equilibrium ground state, the spin
impurity atoms could propagate towards the direction the external
linear potential increases, which means a particle with initial
momentum zero moves upwards in a gravity field in our case. This
unconventional phenomenon, similar to the repulsive bound atom
pairs, is a result of the conservation of the energy. We would also
discuss the profound connection between these two non-equilibrium
phenomena.

First, we propose the experimental set-up of our non-equilibrium
system. It is motivated by a recent experiment\cite{Palzer}, where
the quantum transport of spin impurity atoms through a
Tonks-Giradeau gas is studied. The departure point of our
experimental set-up is loading the $^{87}Rb$ atoms with the
hyperfine ground state $|F=1,m_F=-1\rangle$ into a 1D optical
lattice, which is along z-direction. We assume that the interaction
between these $^{87}Rb$ atoms is very large and the filling factor
is one, thus the ground state of the system is the Mott
state\cite{Stoferle}. An additional harmonic magnetic trap is
implemented to provide a vertical confinement for the atoms and
prevent the atoms escaping from the optical lattice due to the
gravity. The harmonic magnetic trap couples with the atoms via the
magnetic dipole interaction, therefore the vertical confinement is
purely magnetic.

\begin{figure}
  \centering
  \includegraphics[width=4cm]{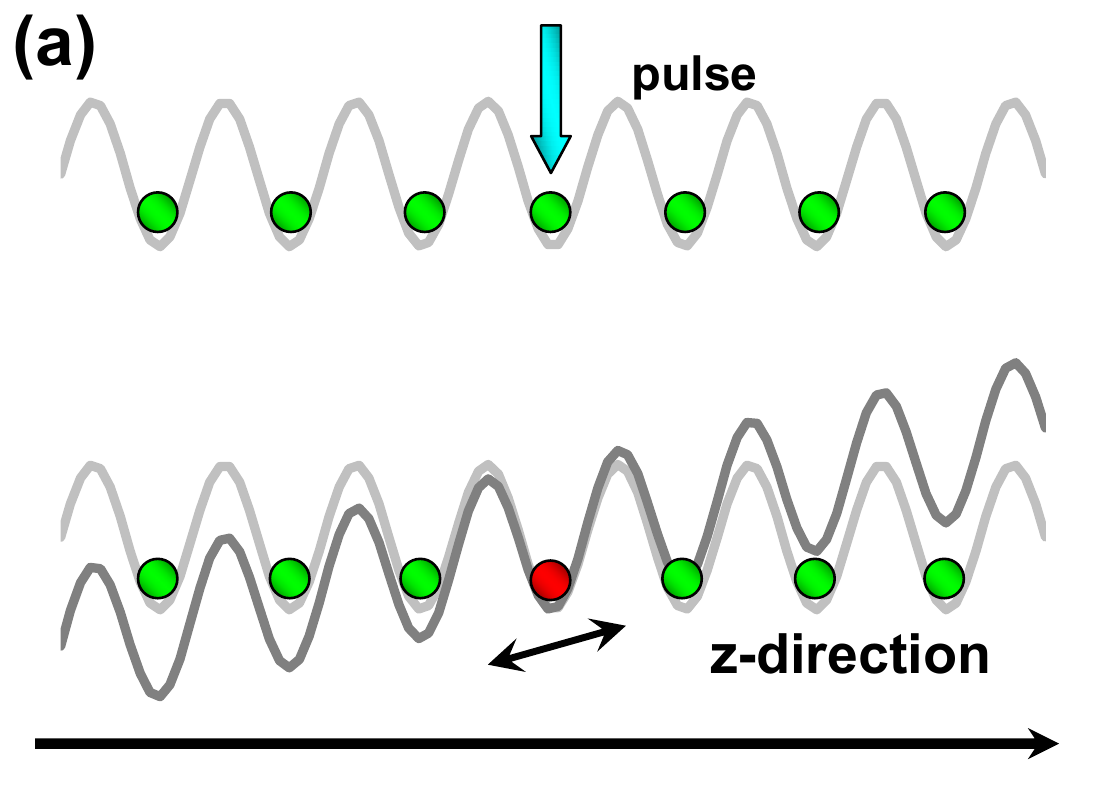}
  \includegraphics[width=4.2cm]{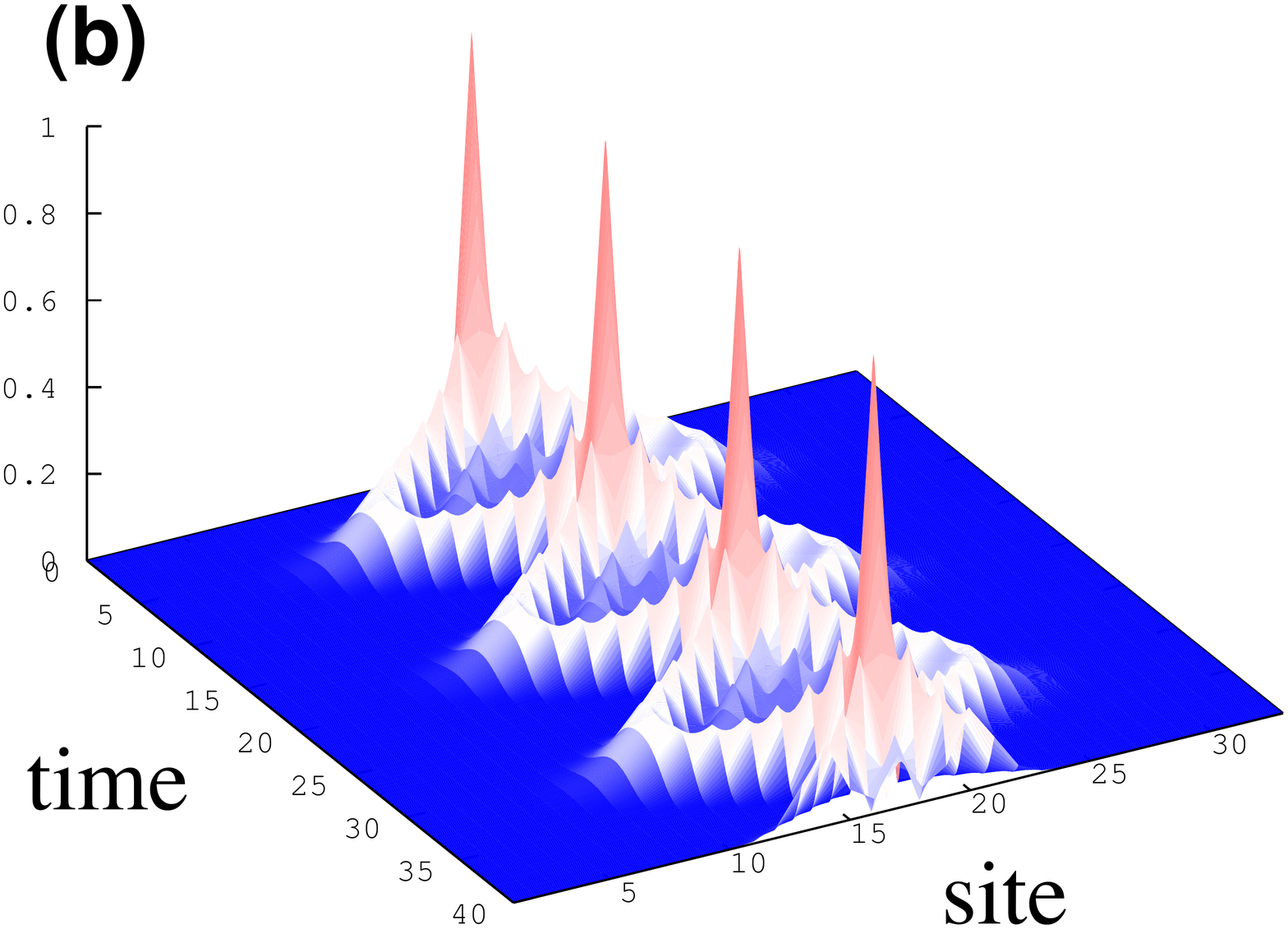}

  \vspace{0.5cm}%
  \includegraphics[width=4.2cm]{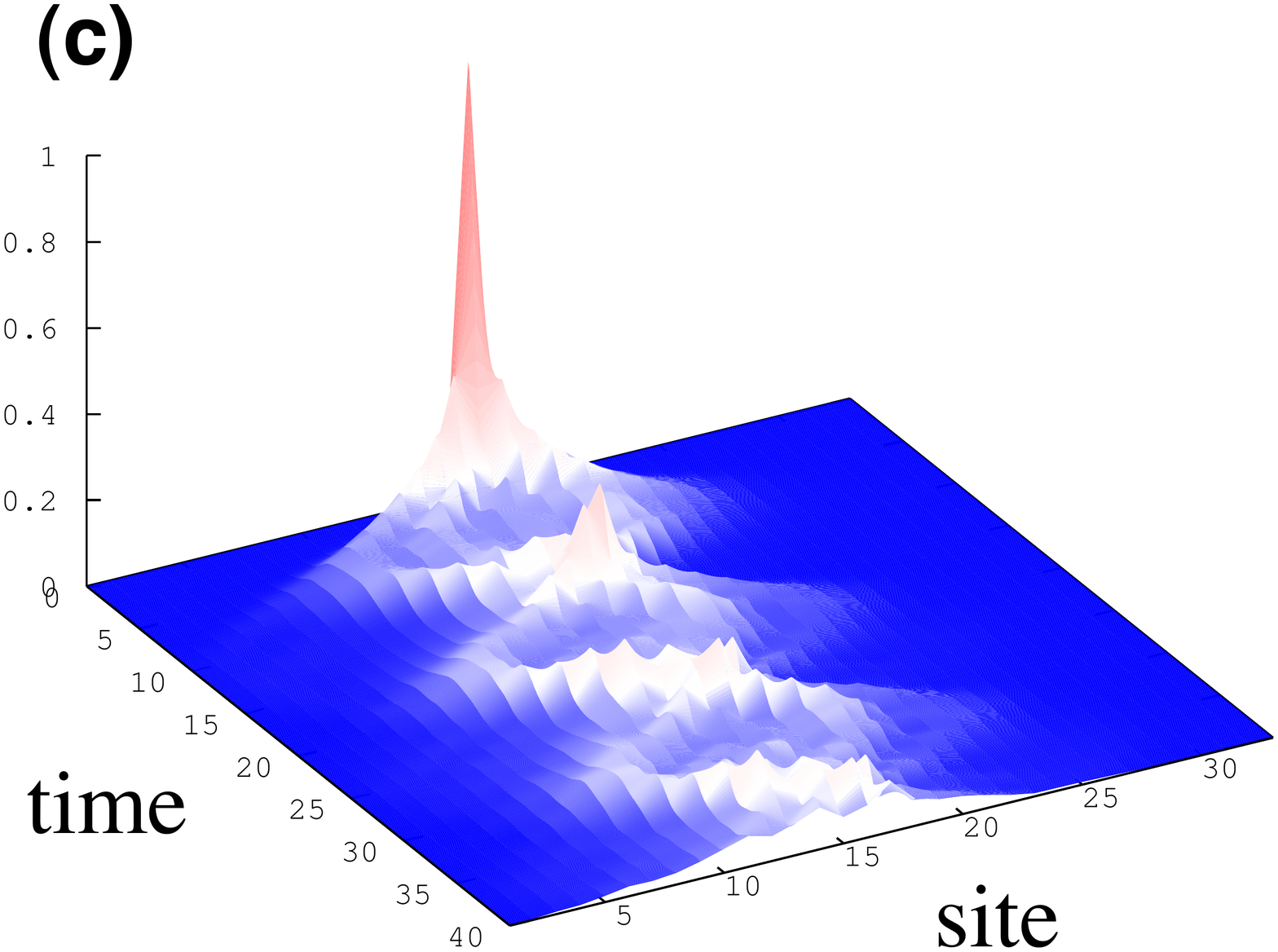}
   \includegraphics[width=4.2cm]{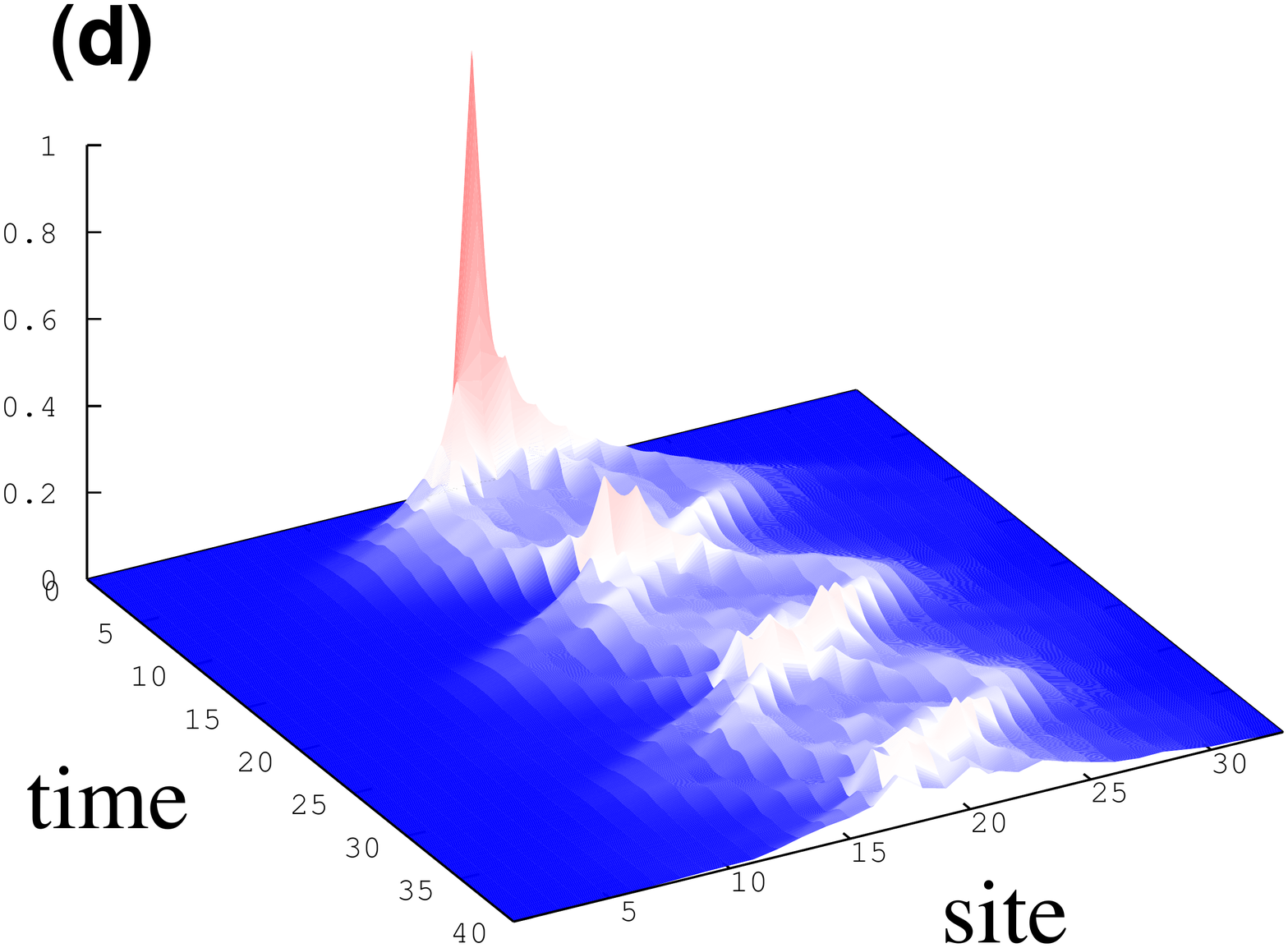}
\caption{(Color online) (a). Experimental setup.
The initial state is prepared by applying rf pulse on single site,
making boson atom transitions from $|F = 1,m_{F} = −1\rangle $ to $|F = 1,m_{F} = 0\rangle $ level. Bosonic atom in the latter level experiences gravity potential. (b).-(d). Evolution of density distribution of the spin impurity, for $G/t=0.5, U/t=0,2,-2$ respectively. }\label{fig:lattice}
\label{fig:occ}
\end{figure}

To prepare the initial state, one pumps the Mott ground state into a
non-equilibrium state by applying a pulse of radio frequency
resonant with the energy gap between the two hyperfine spin states
of $^{87}Rb$ atom: $|F= 1, m_F=-1\rangle$ and $|F=1, m_F=0\rangle$
and inducing a transition $|F= 1, m_F=-1\rangle\rightarrow|F=1,
m_F=0\rangle$\cite{Esslinger}. Below we denote the $|F= 1,
m_F=-1\rangle$ as trap atoms and $|F=1, m_F=0\rangle$ as impurity
atom.  The spatial width and position of the pulse could be
controlled experimentally\cite{Palzer}. In our case, we constraint
the pulse to only induce the transition  on one site and  produces
one impurity, as shown in Fig.1 (general case containing more than
one impurities will be discussed below). Notice the magnetic moment
of the impurity atom $|F=1, m_F=0\rangle$ is zero thus it is not
confined by the harmonic magnetic trap and experiences an external
linear potential produced by gravity. The experimental set-up
proposed here is different from that in Ref.\cite{Palzer} in the way
that in our case, the bosons are loaded in a 1D optical lattice
along the z-direction, rather than the Tonks-Girardeau gas loading
in a continuous 1D space. Without the constraint of the periodic
potential along the z-direction, an impurity atom in a gravity field
would be accelerated downwards and finally escape from the system,
as shown in Ref.\cite{Palzer}. In our case, however, the dynamics of
the impurity atom becomes complex and nontrivial due to the optical
lattice structure as well as the interaction between the impurity
atom and the trap atoms. Without the interaction, the single
particle dynamics is known as Bloch oscillation\cite{Blochf}, which
has been observed in ultracold atomic
systems\cite{BenDahan,Wikinson}.

\begin{figure}
[tbp] \centering
   \includegraphics[height=7cm, width=8cm]{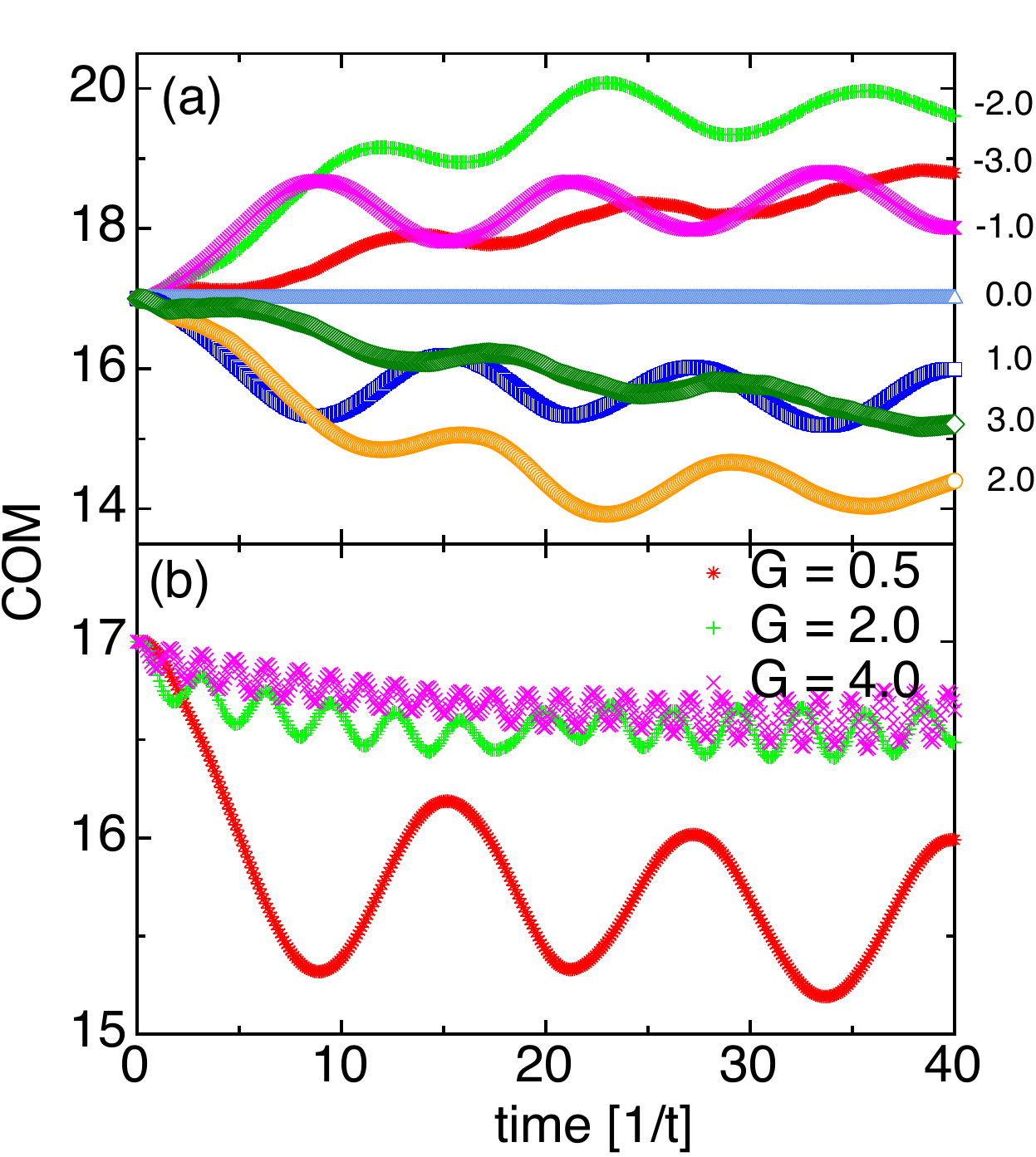}
  \caption{(Color online) The COM of the spin impurity for (a). $G/t=0.5$ with different interaction strengths, the $U/t$ values are indicated at right axis. (b).  $U/t=1$ with different gravity fields, which could be attained by changing the lattice spacing or choosing different type of trapped atoms.} \label{fig:varUandG}
\end{figure}

The Hamiltonian in this system can be described by a 1D
two-component Bose-Hubbard model in a linear external potential:
\begin{equation}
H=-t\sum_i (b_i^\dag b_{i+1}+a_i^\dag a_{i+1}+ h.c)+U\sum_i
n_i^a n_i^b+G\sum_i in_i^a
\end{equation}
where $a_i$($b_i$) is the annihilate operator for the impurity
(trap) atoms. $U_{ab}=U$ is the interspecies on-site interaction
strength. We assume the intraspecies on-site interaction $U_{aa}$
and $U_{bb}$ are repulsive much larger than $U_{ab}$, thus the boson
could be considered as hard-core for the bosons with same species.
$G=mga_0$ is the potential gradient of the gravity, where m is the
mass of  $^{87}Rb$ atom and $a_0$ is the lattice constant of the
optical lattice. The energy scale of $G$ could be estimate as $\mu
K$, which is in the same order of magnitude of $U$ and $t$. We
assume that the energy scale of the gap between the s-band and
p-band is much larger than that of the parameters $(t,U,G)$ in our
Hamiltonian.(1), therefore all the atoms are confined within the
s-band and no interband physics involved. As we analyzed above, only
the impurity atoms experience the gravity potential while the trap
atoms are still confined by the magnetic trap. The non-equilibrium
initial state can be represented as:
$|\uparrow\uparrow\cdot\cdot\cdot\downarrow\cdot\cdot\cdot\uparrow\uparrow\rangle$,
where $\uparrow$ ($\downarrow$) denotes the trap (impurity) atoms.
The impurity atom is initially localized in the center of the
optical lattice, while all the trap atoms are still in the Mott
phase. Obviously, this initial state is not a eigenstate of the
Hamiltonian.(1) thus it would evolve. The time evolution of the
system  is controlled by the parameters in Hamiltonian.(1). Below we
would study the dynamics of the impurity atom and show how it
depends on the parameters ($U/t, G/t$) in Hamiltonian.(1) using the
TEBD method\cite{Vidal}. Total number conservation is used to reduce
the computational effort. A straightforward treatment of the two
component HCB needs local dimension $d=4$, to reduce the
computational effort, unfold technique is used and thus the local
dimension is reduced to $d=2$. The pay off is that the chain length
increases to $2L$ and next-nearest neighbor interactions are
introduced. Swap technique \cite{Vidal,Danshita} is used to deal
with next-nearest neighbor interaction within TEBD method. Both the
unfolded ($d=2$) and folded ($d=4$) algorithm are implemented and to
cross check the results. In this work we deal with chain length
$L=33$, and the spin impurity was create at the $17th$ site. Finite
size effect has little effect to the conclusion since the Bloch
oscillation physics confines the atom near its original position. In
the course of real time evolution we take the truncation dimension
$\chi=80$ and time step $\Delta\tau=0.05$, the convergence is
checked by taking larger $\chi$.

The time evolution of density distribution of the impurity atom is
shown in Fig.1,(b)-(d).  The dependence of the motion of impurity
atom's center of mass (COM) on the interaction (U) and potential
gradient (G) is shown in Fig.(2).  If there is no interaction (U=0),
the COM of the impurity atom does not move at all. This phenomenon
could be understood from the dynamics of a single particle under a
combination of periodic and a linear external potential. The presence of the
external potential breaks the translational symmetry, render momentum $k$ not a
good quantum number. However, by choosing
the temporal gauge (the scalar potential $\phi=0$), the external potential is taken into account by
a time-dependent gauge field $A(t)=Gt$ and
the momentum $k$ is replaced by $k-A(t)$. The noninteracting
Hamiltonian can be rewritten in the momentum space as : $
H(t)=\sum_k\epsilon_{k-A(t)}a^\dag_k a_k$, where $k-A(t)$ is good quantum number. Suppose initially the
particle locates on the point $x=0$, the initial state can be
represented in the momentum space as:
$|x=0\rangle_{t=0}=\frac1L\sum_k|k\rangle$ ($e^{ikx}$=1 because
$x=0$), where $|k\rangle$ is the single particle eigenstate in the
momentum space and the sum is over the first Brillouin zone. In the
process of time evolution, all the $|k\rangle$ modes experience
Bloch oscillation from different initial momentum. The localization of
the COM is actually a consequence of the superposition of all the
momentum modes $|k\rangle$ in  our initial state, the Bloch
oscillation in $|k\rangle$ and $|-k\rangle$ modes cancel with each
other and the net effect is zero.

The single particle picture does not work when $U\neq 0$, thus we
deal with the time evolution by TEBD. If $U>0$, the COM of the
impurity atom would move downwards until it reaches a steady
quasi-equilibrium position, while interesting phenomenon emerges for
$U<0$. The COM of the impurity atom would move upwards though it is
experiencing a gravity potential. We also find that the dependence
of the motion of COM on $U$ is not monotone. Notice in both case of
large U and small U, the COM is always constrained around the
original position. The maximum deviation of the COM away from
original point occurs when $U/t=\pm2$. The dependence of the dynamic
of the COM on the gravity gradient G is also nontrivial. We notice
that for larger G, the COM of the impurity would drop to a higher a
final quasi-equilibrium position. The period of the oscillation is
approximately the Bloch period $2\pi/G$.

To understand the origin of the counter-intuitive phenomenon, we
perform a Jordan-Wigner transformation to map our system into a
interacting fermionic system (considering the hard-core nature for
the intraspecies boson). Then we use the particle-hole
transformation on the trap atoms, thus the problem could be
translated into a two-particle problem due to the half-filling
condition in our initial state. The initial configuration in
Fig.\ref{fig:lattice}(a) could be considered as two particles (an
impurity atom and a trap hole) located on the same site (doublon) of
the lattice. Below we mainly focus on the attractive condition
($U<0$) in original model, which could be mapped into a two-particle
problem with replusive interaction ($\tilde{U}>0$) under the
particle-hole transformation. If $\tilde{U}\gg t$, the dynamics of
this two-particle problem reminds us of the physics in the the
repulsively bound atom pairs\cite{Winkler, Daley}. The initial state
is a doublon state with a high energy $\tilde{U}$.  If the two
particles are separated, the loss of the interaction energy
($\tilde{U}$) should be compensated by the increasing of the kinetic
energy of the single particle to preserve the total energy of the
system. However,in the periodic lattice structure, the kinetic
energy of a single particle has a maximal value ($2t$). Therefore,
when $\tilde{U}\gg t$, a doublon  would be stabilized by the strong
repulsive interaction, fig.\ref{fig:phasediag}(c). In our case,
interesting thing happens for smaller U, where the gravity field
play a key role.  It is possible that the loss of the interaction
energy could be compensated by the increasing of the gravity
potential energy of the impurity atom to preserve the total energy,
as shown in fig.\ref{fig:phasediag}(b). The energy transfer between
the interaction energy and gravity potential energy results in the
anomalous  dynamics of the impurity atom.

\begin{figure}
    [tbp] \centering
\includegraphics[height=7cm, width=8cm]{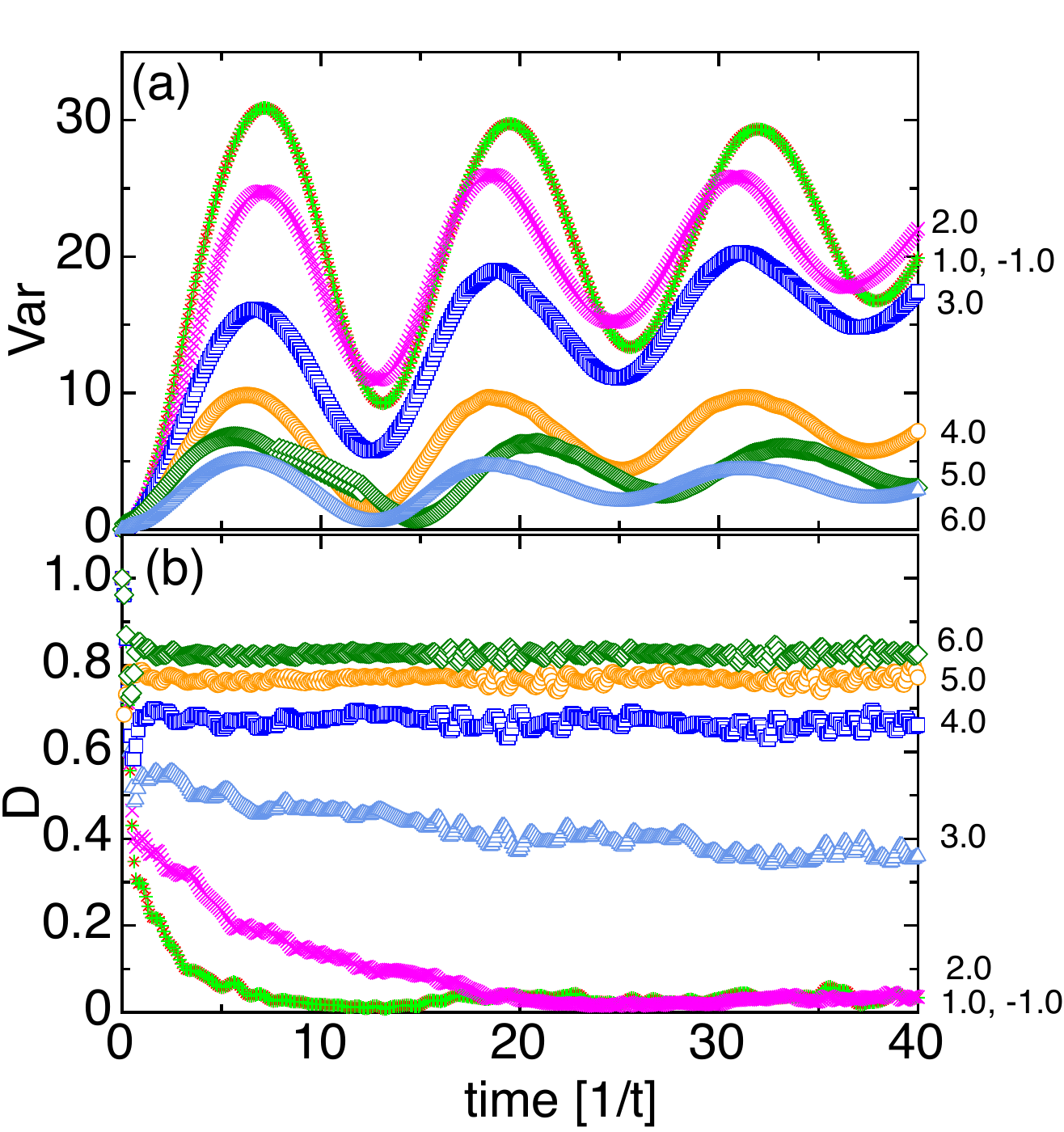}
\caption{(Color online) (a). Variance of the COM of spin down boson,
(b). Total doublon numbers in the lattice for $G/t=0.5$ and different interacting strength, $U/t$ values are indicated at right axis.} \label{fig:var_dbocc}
\end{figure}

To verify this qualitative picture, we again use TEBD to study the
two particle dynamics from the given initial doublon state. The
result is shown in Fig.\ref{fig:var_dbocc},  We introduce the
variance of the COM of the impurity atom, defined as
$Var=\langle\sum_i i^2 n_{i}^a\rangle- \langle(\sum_i
i~n_{i}^a)^2\rangle$, to characterize the width of the wave packet,
and the total doubly occupancy number $D=\langle\sum_i n_{i}^a
\widetilde{n}_{i}^b \rangle$ to study the fate of the initial
doublon state, where $\widetilde{n}_{i}^b$ denotes the number of the
trap hole on site $i$. The variance decrease as the increase of the
interacting strength, although it shows oscillation which is given
by the Bloch oscillation. This means that larger interacting
strength squeeze the wave packet harder since dispersion of it lose
more energy. For very large $\tilde{U}$ (compared to $G$ and $t$),
$Var$ is approaching to $0$ and $D$ to $1$, which means the doublon
state is stabilized by the strong repulsive interaction. In
addition, the doublon is localized on its original position since
motion of it towards any direction violate the energy conservation
in the gravity field.  For smaller $\tilde{U}$, interacting energy
could be released to the energy continuum by pumping the impurity
atom upwards to gain potential energy. This picture is further
confirmed by the rapid decreasing of the total double occupancy
number, which means the doublon dissolve shortly for small
interacting strength. The COM and its variance of the impurity atom
could be measured by time resolved tomographic measurement of the
impurity density distribution \cite{Palzer}.

\begin{figure}[!t]
\includegraphics[width=8.5cm] {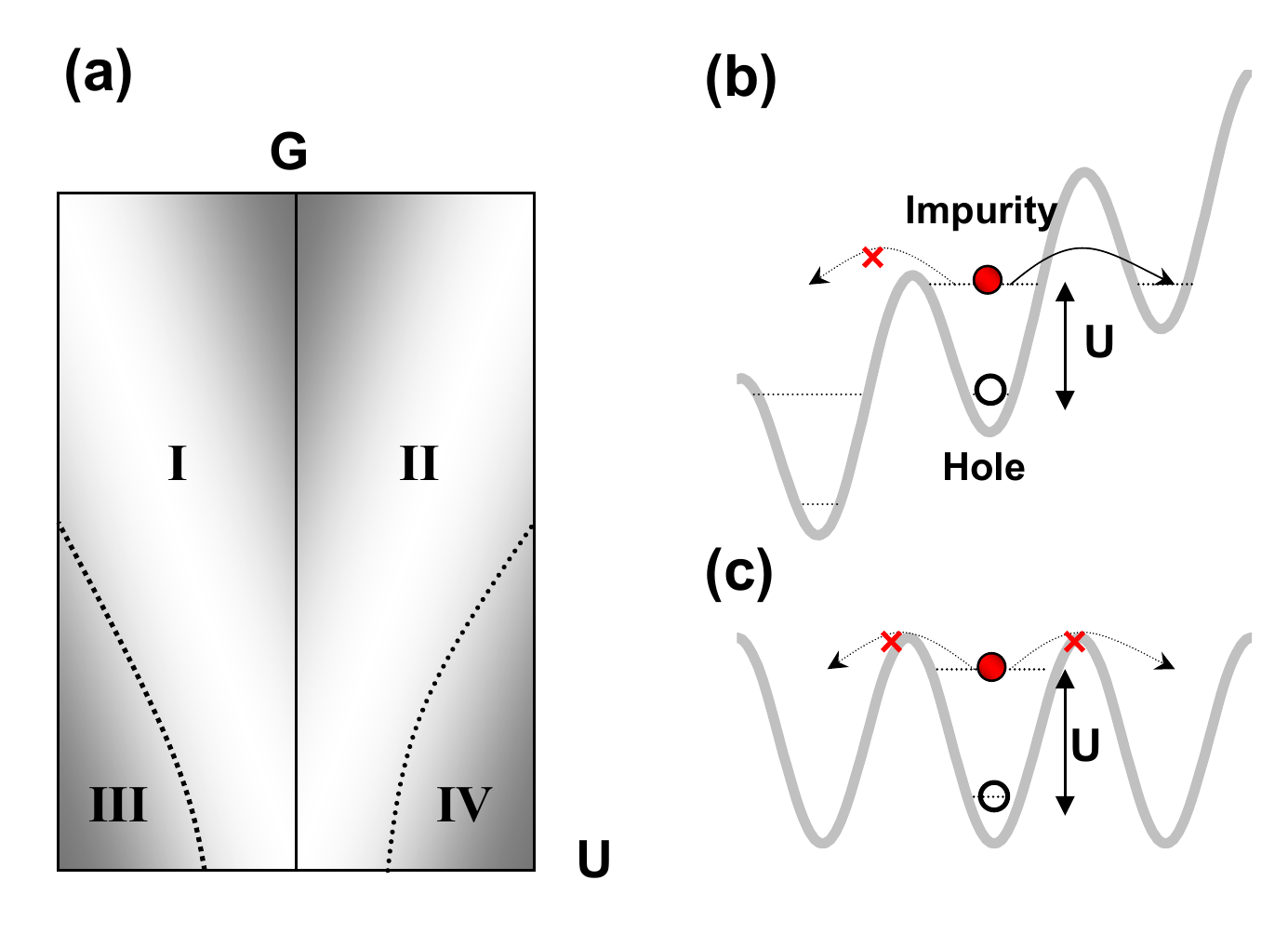} \caption{(Color online)
(a). Sketch phase diagram of
the final quasi-equilibrium state of the spin impurity: The impurity atom moving
(I) upwards or (II) downwards for intermediate attractive or repulsive interaction. Trapped exciton
state induced by (III) strong attractive (IV) repulsive interaction.
(b). Schematic view of impurity atom moving upwards to preserve the total
energy. (c). Exciton be trapped by strong attractive interaction where $|U|\gg G$. }
\label{fig:phasediag}
\end{figure}

Translate back into spin impurity language, the impurity atom and a
trap hole bind together  to form an exciton (which is the doublon in
the two-particle picture) for both large attractive and repulsive
interaction. For the noninteracting case, the exciton dissolves
immediately and Bloch oscillation physics dominates.  For the
intermediate interacting strength, the impurity atom will move
upwards or downwards depending on the sign of the interaction. The
sketch phase diagram is shown in Fig.\ref{fig:phasediag}. Notice
that there is crossover instead of phase transition between
different regions. The general principle guiding the time evolution
of the strongly interacting systems is that the system always
evolves toward a maximum entropy state with the constraint of the
conservation of good quantum numbers (such as energy and total
particle number)\cite{Rigol2,Cramer}. So the competition between the
energy and entropy play a key role in determining the properties in
the final quasi-equilibrium steady state.

Finally, we would discuss the dynamic of the impurity atom evolving
from two different initial configurations, we focus on the
attractive condition to show the dependence of the anomalous
phenomenon on the initial configuration we choose. The first case is
that the filling factor in the initial state is not one, which means
the impurity would propagate in an environment of the
quasi-superfluid of the trap atoms. The attractive interaction
between the impurity and the BEC atoms leads to a
polaron\cite{Bruderer}, and its transport properties have been
studied previously\cite{Klein}. We choose the initial state as a
quasi-superfluid with 31 hard-core bosons filled in the 1d lattice
with 33 sites. We can find that the COM of the impurity atom does
not move upwards, different from that in the Mott initial state.
This phenomenon can also be understood by the particle-hole
transformation on the trap atoms. In this case, the trap hole is not
localized in the initial configuration due to the
quasi-superfluidity of the trap atoms, which means there is no
high-energy doublon initial state. The energy transfer picture is no
longer available, thus the anomalous  transport phenomenon
disappears. Actually, we can always prepare a Mott plateau region in
lattice by adding an external harmonic trap potential, even the
filling factor is not one, and the "Negative Mass" phenomena will
still appear. The other configuration is to apply a pulse of radio
frequency on more than one site and produce many impurity atoms in
the initial configuration. The initial state is chosen as five
impurity atoms (from site 15 to site 19) located in the middle of
the lattice while all other site is filled with trap atoms. We can
find in this case the COM can still move upwards in the attractive
interaction and the main result of this paper  is not qualitatively
changed by this multi-impurity effect.

\section{Acknowledgment}
The work is supported by NSFC. the Knowledge Innovation Project of
CAS, the National Program for Basic Research of MOST. XCX is
supported by US-DOE and US-NSF. LW thank Xi Dai and Yuan Wan for helpful discussions.
We thank X.-M. Cai for bring Ref\cite{Palzer} into our attention.

\end{document}